\documentclass{webofc}
\usepackage[varg]{txfonts}   
\def\pp{pp\xspace}

\def\pA{p$A$\xspace}

\def\AA{$AA$\xspace}
\def\PbPb{Pb-Pb\xspace}

\newcommand{\pythia}{P\protect\scalebox{0.8}{YTHIA}\xspace}
\newcommand{\rivet}{R\protect\scalebox{0.8}{IVET}\xspace}

\begin{document}
\title{Sources of multiparticle correlations}
%
%
\subtitle{-- a microscopic perspective}

\author{\firstname{Christian} \lastname{Bierlich}\inst{1,2}\fnsep\thanks{\email{christian.bierlich@thep.lu.se}}
}

\institute{Department of Astronomy and Theoretical Physics, Lund University, Sölvegatan 14A, Lund, Sweden.
\and
          Niels Bohr Institute, University of Copenhagen, Blegdamsvej 17, Copenhagen, Denmark.
          }

\abstract{%
Multiparticle correlations is a hallmark measurement characterizing the behaviour of the assumed Quark Gluon Plasma in heavy ion collisions. In these proceedings an alternative, microscopic approach is presented, based on interacting strings and multiparton interactions.
	}
\maketitle
\section{Multiparton interactions: From \pp to \AA}
\label{sec:mpi}
\begin{figure}[h]
	\includegraphics[width=0.5\linewidth]{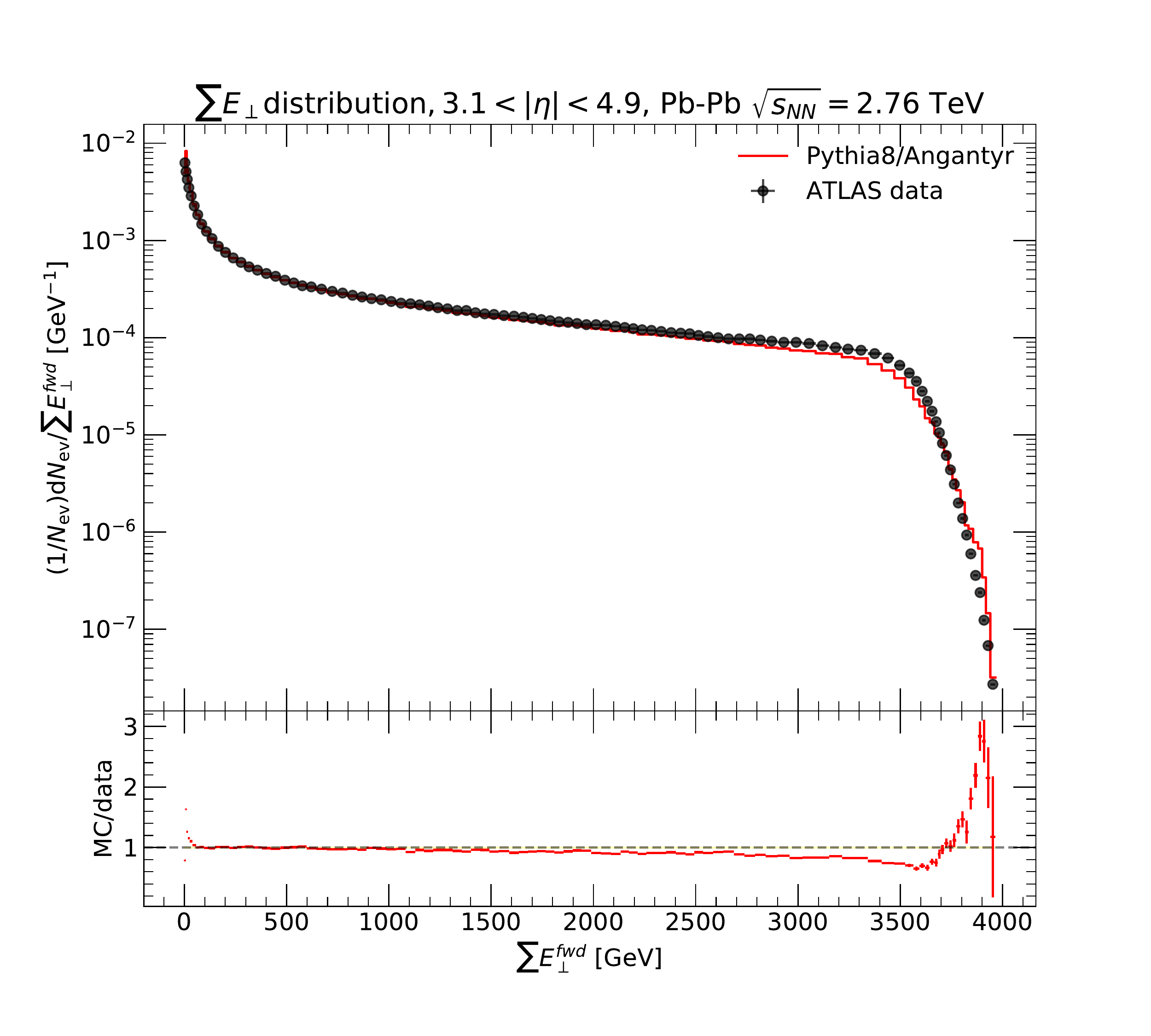}
	\includegraphics[width=0.5\linewidth]{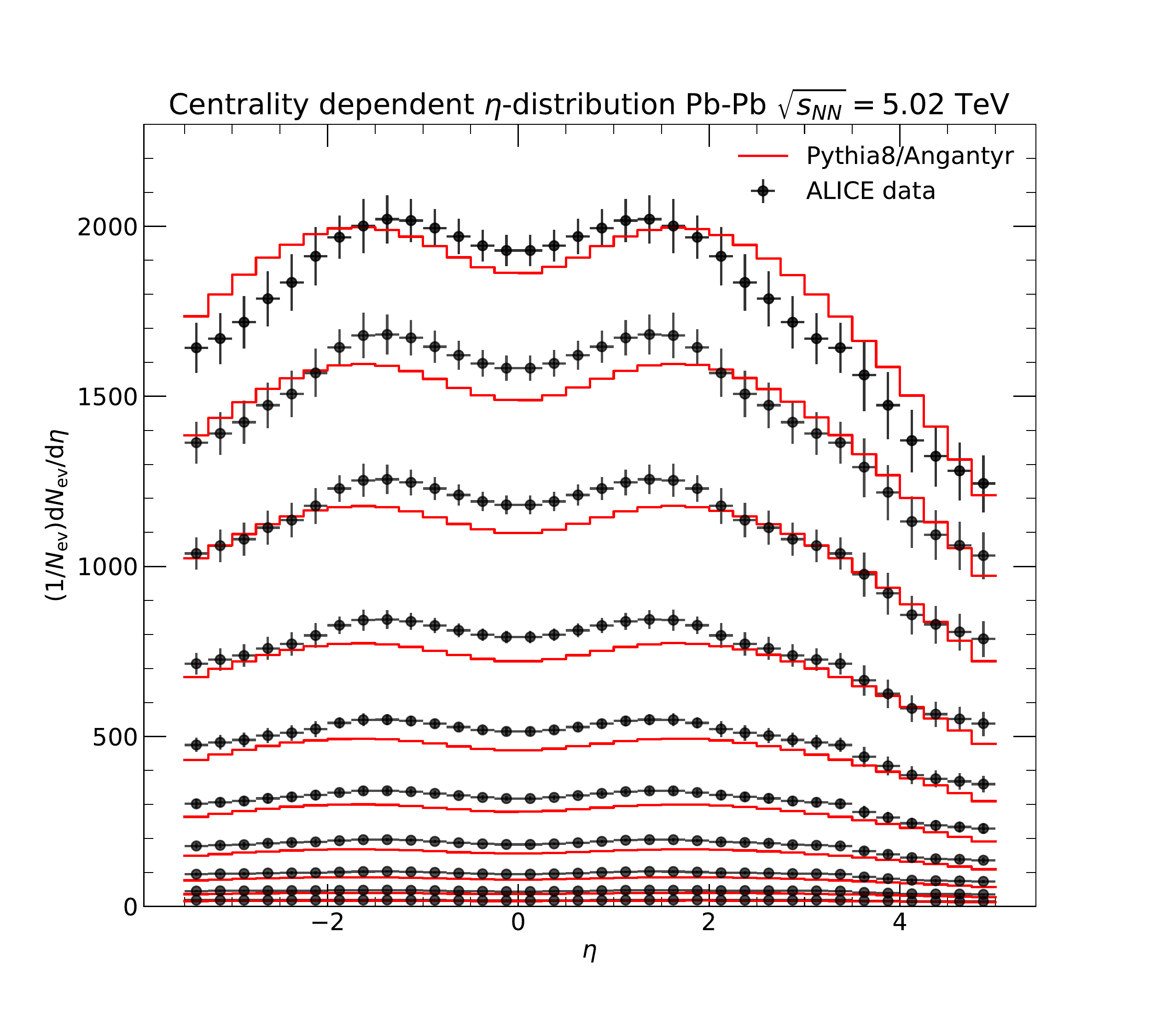}
	\caption{\label{fig:multiplicity}Forward energy deposition (left) and multiplicity as a function of pseudo-rapidity (right) in \PbPb collisions shown, compared to simulation by \pythia+Angantyr.}
\end{figure}

In high energy collisions of protons, the presence of multiparton interactions (MPIs) are ubiquitous, and to a large extent successfully
modelled by considering individual interactions as almost independent\footnote{Not entirely independent, as parton density corresponding to the extracted parton is rescaled by a factor $1-x$, and energy--momentum conservation is obeyed.} perturbative scatterings. This model was introduced
to the \pythia event generator \cite{Sjostrand:2014zea} by Sj\"ostrand and van Zijl \cite{Sjostrand:1987su}, and takes as a starting point the $2\rightarrow 2$ scattering cross section, with its low $p_\perp$ divergence regularised using a parameter $p_{\perp0}$:
\begin{equation}
\label{eq:mpi-model}
\frac{\mathrm{d}\sigma_{2\rightarrow 2}}{\mathrm{d}p_\perp^2} \propto
  \frac{\alpha_s^2(p_\perp^2)}{p_\perp^4} \rightarrow
  \frac{\alpha_s^2(p_\perp^2 + p_{\perp 0}^2)}{(p_\perp^2 + p_{\perp 0}^2)^2}.
\end{equation}
Naively, this adds sub-scatterings, which in turn each draws two strings (for gluonic interactions) from projectile to target, filling the whole available rapidity range with final state particles. As it was already realized in the original paper, such a simplistic treatment fails to describe data. In particular the rise of $\langle p_\perp \rangle$ with $N_{ch}$ cannot be accounted for. The resolution was to colour-reconnect soft partonic scatterings to hard ones, with the guiding principle of minimizing the total potential energy in the string. Such a treatment can correctly reproduce $\langle p_\perp \rangle (N_{ch})$, and it was later shown to also give effects similar to collectivity \cite{Ortiz:2013yxa}, also reproduced in later, more elaborate models \cite{Bierlich:2015rha}. These collective effects are, however, all short range in rapidity \cite{Bierlich:2018lbp}.
The challenge of extending the MPI formalism to \pA and \AA, lies not in scaling the uncorrelated partonic collisions to a collision systems involving collisions of several nucleons, but rather how the latter part should be superimposed. In refs. \cite{Bierlich:2016smv,Bierlich:2018xfw} the Angantyr model, based on the wounded nucleon model \cite{Bialas:1976ed} and Regge theory, was introduced, where instead of applying colour reconnection, absorptive (i.e. inelastic non--diffractive) nucleon--nucleon collisions are ordered in primary and secondary collisions (see ref. \cite{Bierlich:2018xfw} for a full description of the procedure). Primary collisions are treated as normal absorptive \pp collisions, and secondary collisions are treated as a single wounded nucleon which, inspired by the Fritiof model \cite{Andersson:1986gw}, contributes as a single string with a mass distribution $\propto dM^2/M^2$. This drastically alters the number of particles produced in both central and forward regions, and for high energy \PbPb collisions both are in good agreement with data, as seen in figure \ref{fig:multiplicity} (left) for forward production and (right) for central production.

The correct description of forward production, as shown in figure \ref{fig:multiplicity} (left) is crucial, as it allows for comparison to data in bins of centrality defined in the same way as done in experiment -- in particular for \pA collisions, the translation from geometric to measured centrality is not straightforward. This task is further eased by the \rivet framework for data comparison \cite{Bierlich:2019rhm} (with which the above figures are produced), which has recently been updated to include heavy ion functionality \cite{Bierlich:2020wms}, centrality included.

\section{String shoving}
\label{sec:shoving}
As mentioned in the previous section, the particle content of the final state, is governed by strings responsible for the partonic
final state fragmenting to hadrons. In the following, and in \pythia in general, this task is carried out by the Lund model \cite{Andersson:1983ia,Andersson:1983jt,Sjostrand:1984ic}. Before hadronizing, strings may interact with each other -- especially in \AA collisions, or high multiplicity \pp, where the density is large -- by pushing or ``shoving'' each other \cite{Abramovsky:1988zh,Bierlich:2017vhg} or by forming colour multiplets, so--called ropes \cite{Biro:1984cf,Bierlich:2014xba,Bierlich:2015rha}. This section is concerned with the former option. 

\begin{figure}[h]
	\includegraphics[width=0.5\linewidth]{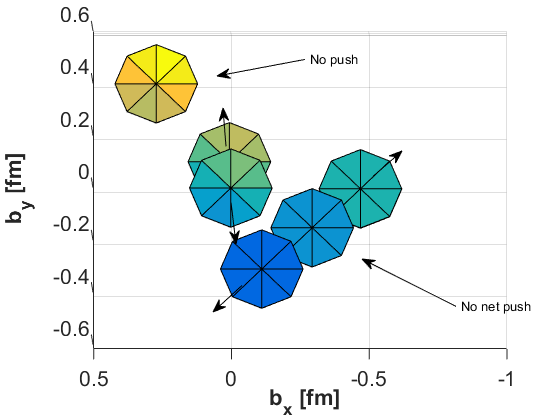}
	\includegraphics[width=0.5\linewidth]{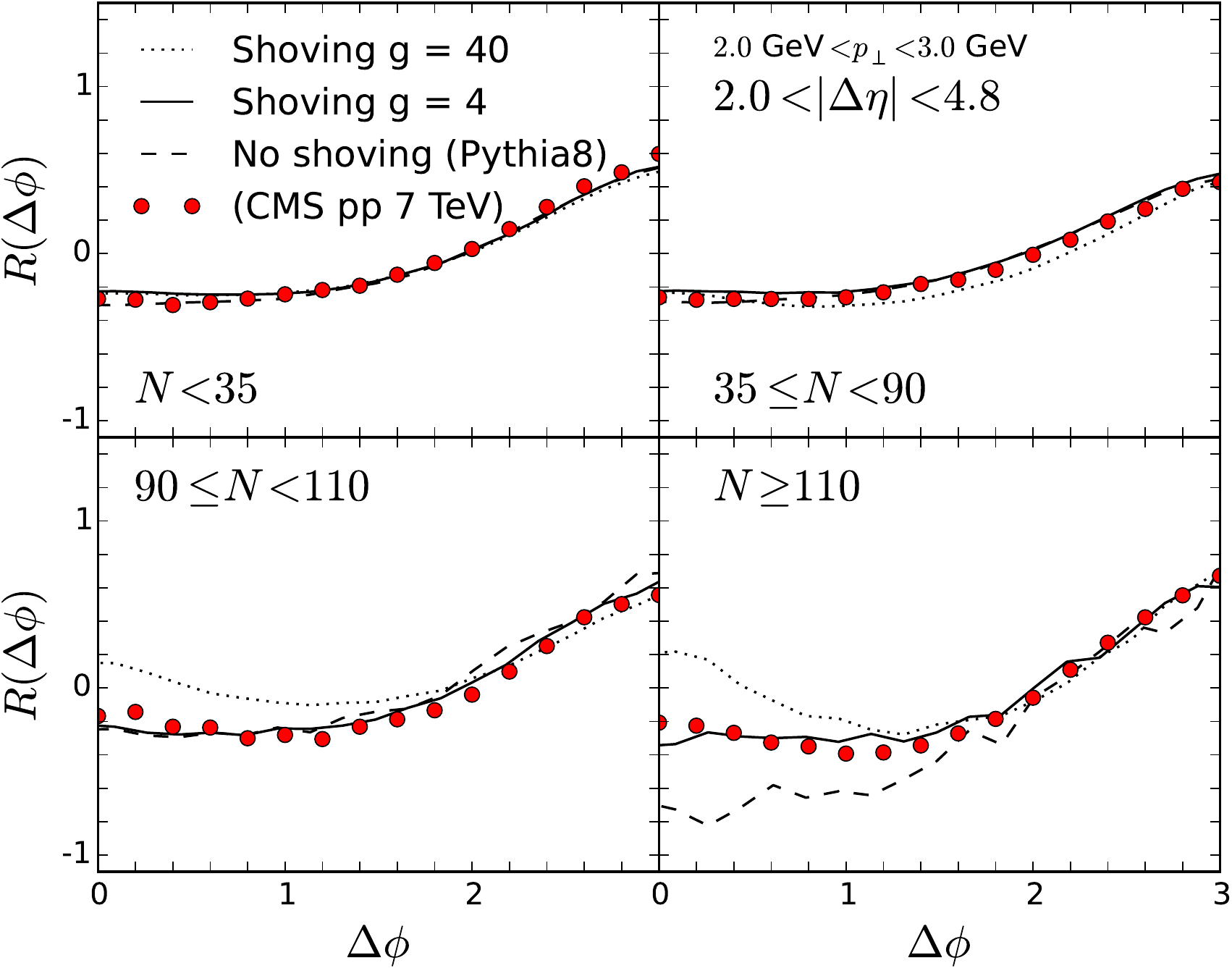}
	\caption{\label{fig:shove-sketch}Sketch of the shoving effect (left) showing how, depending on the spatial distribution of strings, there may or may not be an effect, as well as (right) the successful reproduction of the ridge in \pp collisions by the shoving model.}
\end{figure}

The physical picture of string shoving in transverse space, for a ``slice'' in rapidity, is sketched in figure \ref{fig:shove-sketch} (left). On top left there is a piece not overlapping with any other piece, and it will not experience any interaction. Around $b_x = 0$ two pieces are overlapping, and they will push each other in the direction of the arrows. Finally, in the configuration with three string pieces, the two peripheral pieces will be pushed away, while the central piece will stay stationary as it experiences no net force. The actual expression for the force, is derived a) with inspiration from the duality between the QCD vacuum and an electric superconductor \cite{Baker:1991bc}, and b) lattice calculations of static properties of a string in transverse space.
According to lattice calculations, the longitudinal color-electric field of a string, has nearly a Gaussian shape \cite{Cea:2014uja}:
\begin{equation}
	\label{eq:field}
	E_l(x_\perp) = \frac{\Phi}{2\pi R^2}\exp\left(-\frac{x^2_\perp}{2R^2}\right),
\end{equation}
where $\Phi$ is the flux.
From equation (\ref{eq:field}), the electric field contribution to the string energy density ($\kappa$) can be calculated. Assuming that only the electric field contributes to the string energy density, the potential energy of a two-string configuration with a separation $d_\perp$, can be calculated. The force per unit string length, on one string from another, can then be calculated as the gradient of the potential energy, giving:
\begin{equation}
	f(d_\perp) = \frac{g\kappa}{R^2} \exp\left(-\frac{d^2_\perp}{4R^2}\right),
\end{equation}
where $g$ is a free parameter, determining how much of the energy goes into the magnetic current, and breaking the condensate.

Strings are allowed to shove each other from some initial time after the collision, understood as the time it takes for a string to reach its equilibrium transverse size, until the strings hadronize at around $\langle \tau^2 \rangle = 2$ fm. In figure \ref{fig:shove-sketch} (right) it is shown how the shoving mechanism, as implemented in Pythia 8.2, reproduces the ridge as observed by CMS \cite{Khachatryan:2010gv}.

\begin{figure}[h]
	\includegraphics[width=0.5\linewidth]{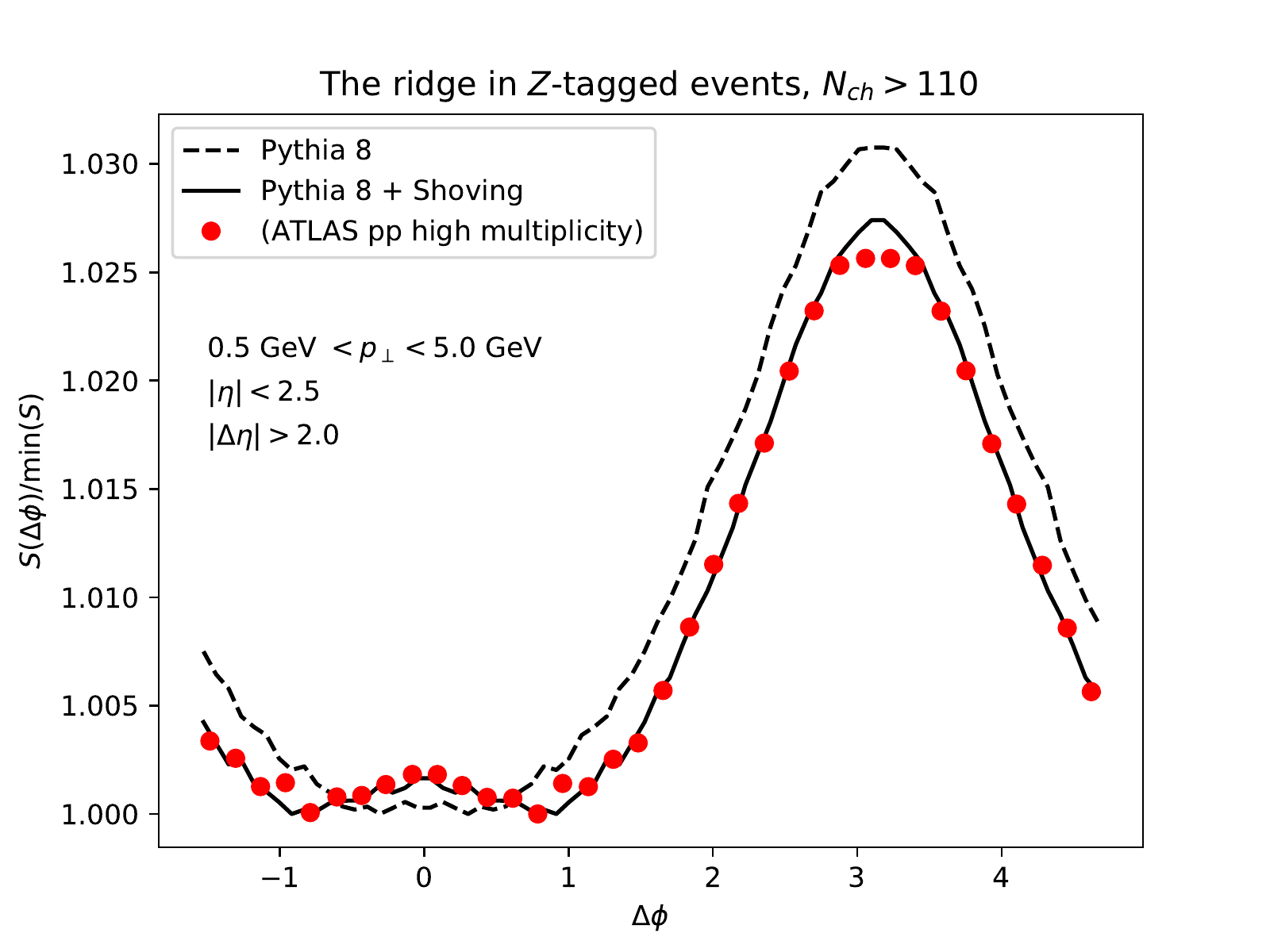}
	\includegraphics[width=0.5\linewidth]{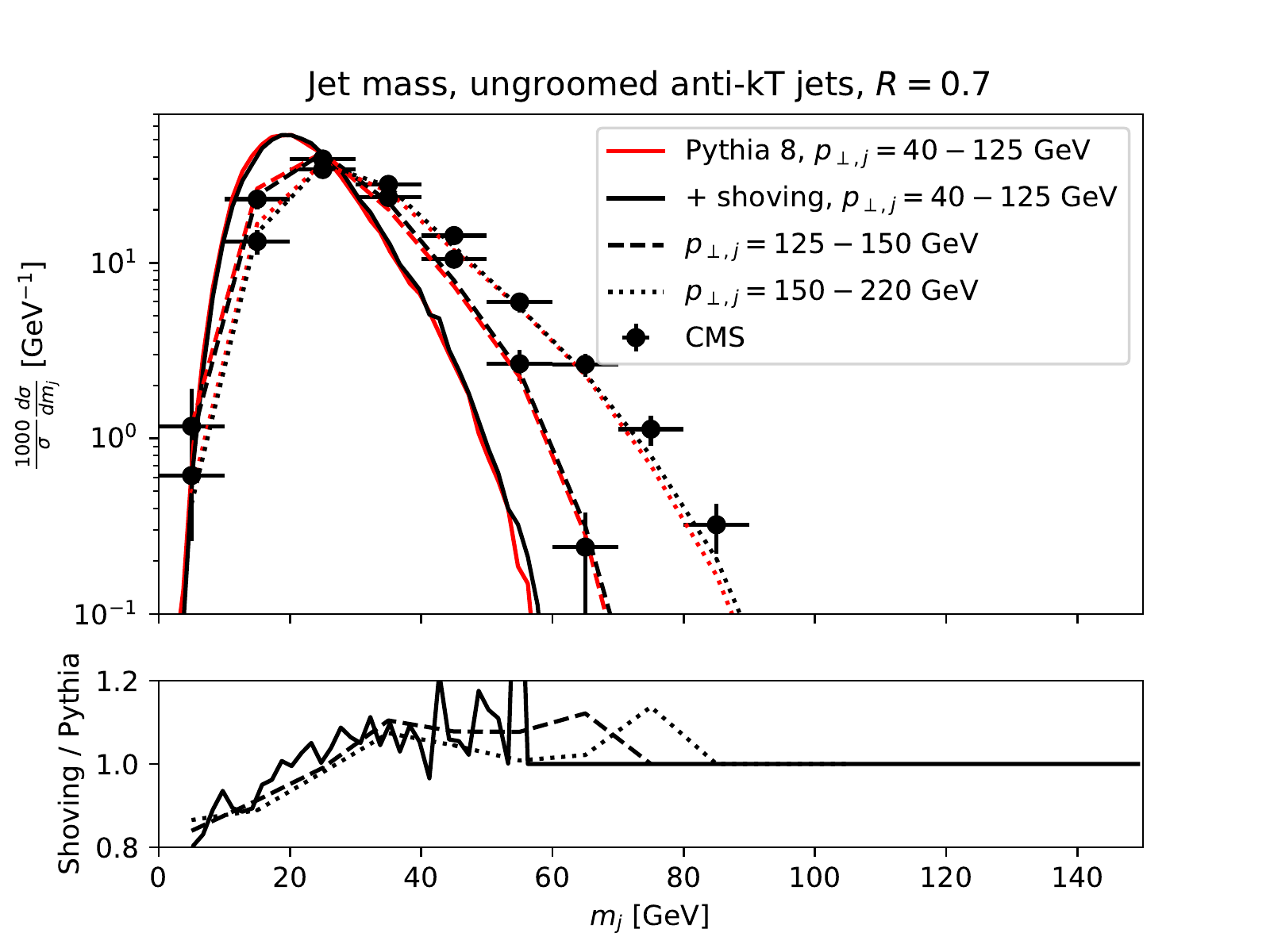}
	\caption{\label{fig:shove-z}The successful reproduction of the ridge in Z-tagged \pp collisions (left) by the shoving model, as well as the effect from shoving on the mass of the recoil jet (right), compared to data.}
\end{figure}

Since the shoving mechanism is purely based on string interactions, it applies equally well for all types of events,
regardless of any type of trigger process. Furthermore, since the string formation is fully integrated in the event generation process, the model allows for full dynamical simulation of kinematical biases present in trigger process events, compared to minimum bias. Consider string interactions in an event with a Z-boson in the di-lepton channel. Since there will not be any string stretched between the leptons, the reconstructed Z will not interact through string interactions, but the rest of the event will. This allows for precision studies of possible jet effects, as the Z gives an unmodified kinematical handle. This possibility was studied in ref.~\cite{Bierlich:2019ixq}, and two key results are shown in figure \ref{fig:shove-z}. On the left side, a comparison to data from ATLAS \cite{ATLAS:2017nkt}, measuring the ridge in Z-tagged events, is shown. It is immediately visible that the ridge effect is reproduced, also around $\Delta \phi = \pi$. Since there is a reasonably hard probe (the Z) present, balancing jets can also be studied. In figure \ref{fig:shove-z} right, the jet mass for anti-$k_t$ jets \cite{Cacciari:2008gp} with $R = 0.7$ in Z-triggered events, is shown with and without shoving enabled, and compared to data from CMS \cite{Chatrchyan:2013vbb}. It is seen that the shoving mechanism affects the jet mass up to about 20\% for low mass jets, pointing to a possible source of jet modifications present in \pp collisions.

\section{The role of the initial state}
\label{sec:initial-state}
\begin{figure}[h]
	\includegraphics[width=0.5\linewidth]{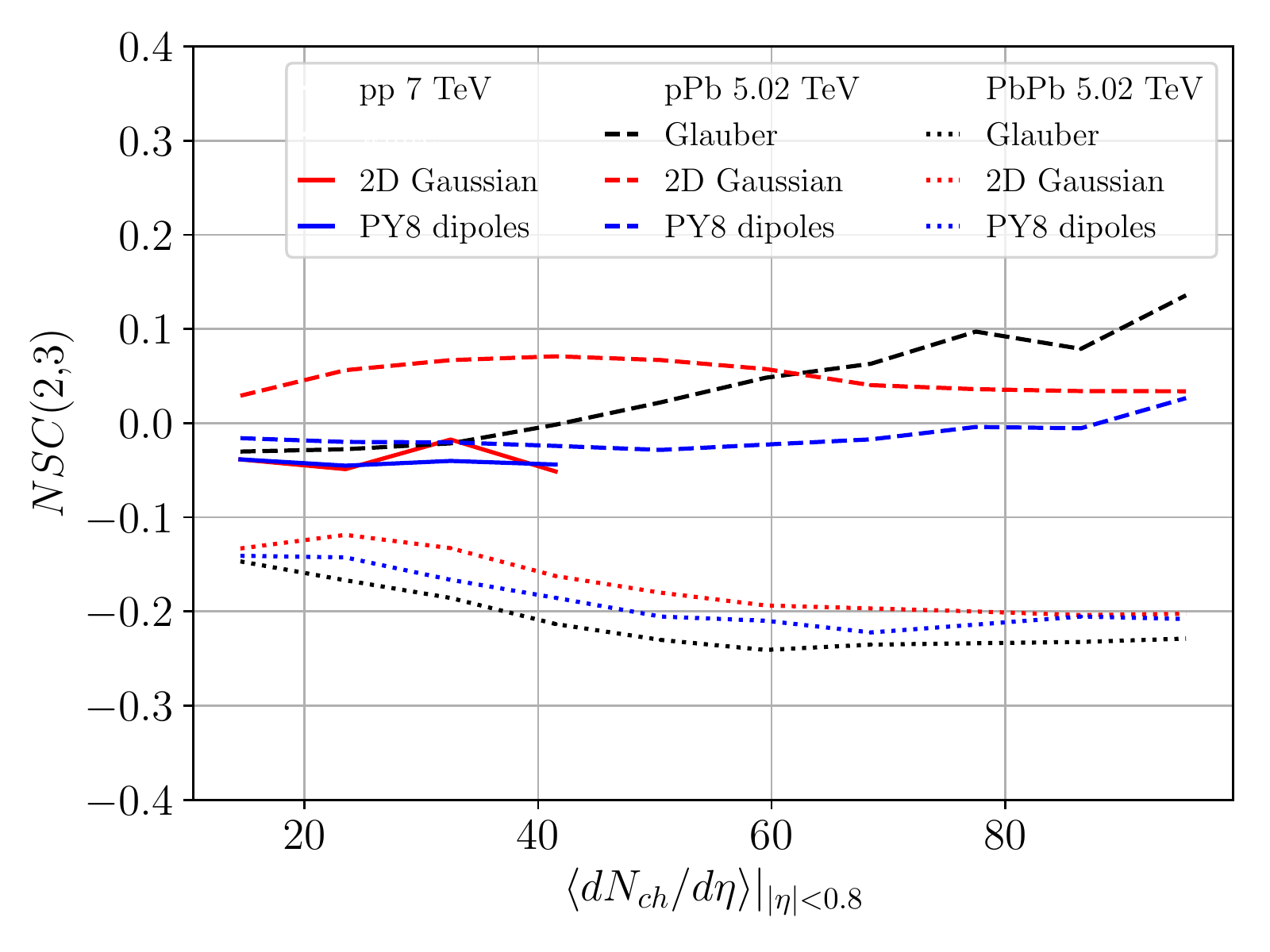}
	\includegraphics[width=0.5\linewidth]{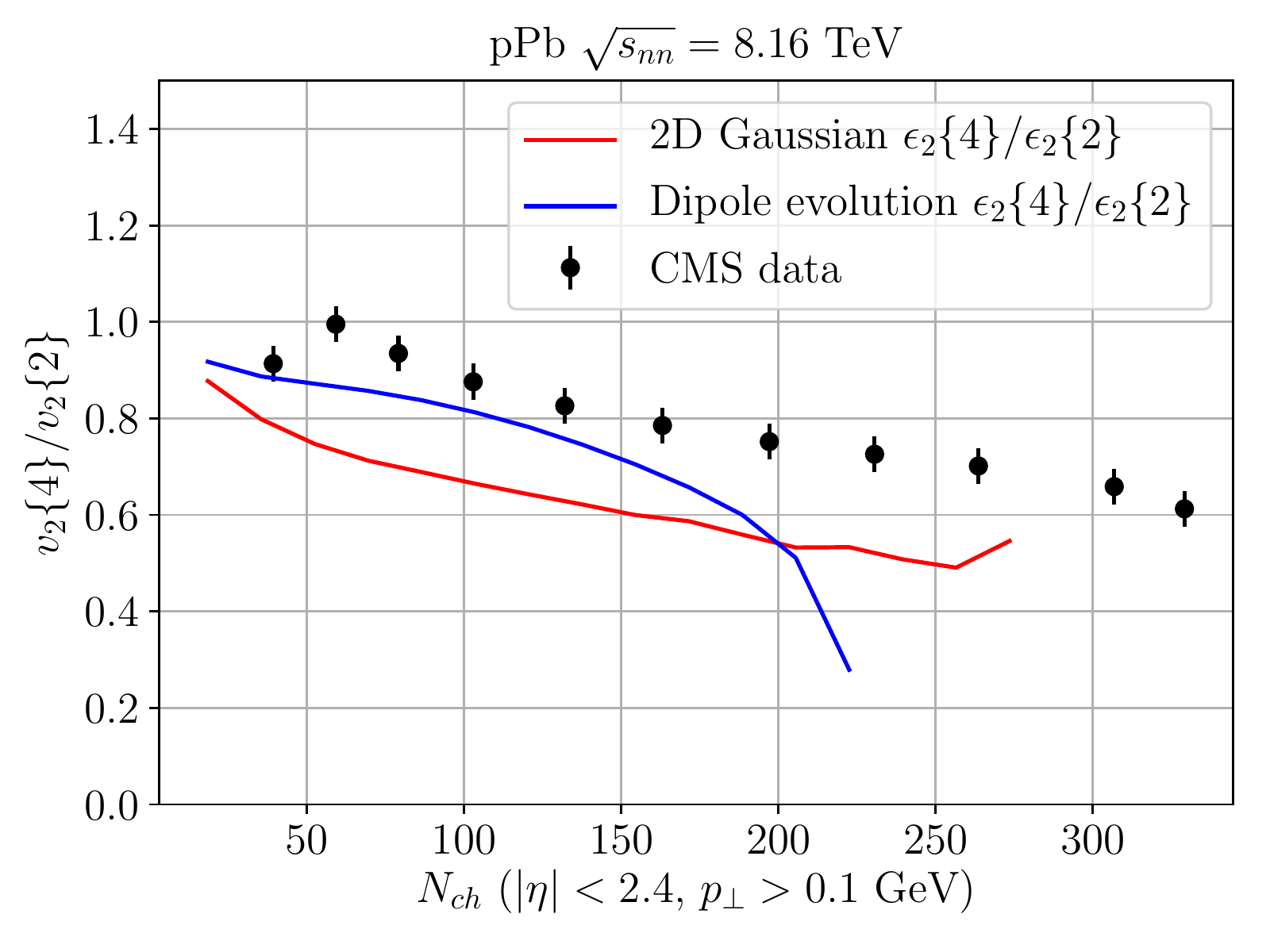}
	\caption{\label{fig:dipole-struc}The normalized symmetric cumulant NSC$(2,3)$ in \pp, \pA and \AA collision by a Glauber treatment, a Gaussian MPI distribution and a Mueller dipole based treatment (left), and a comparison of Gaussian MPI distributions and a dipole based one, compared to data for flow fluctuations in \pA collisions (right).}
\end{figure}

It is a well known feature of hydrodynamic treatments of heavy ion collisions, that flow coefficients such as $v_2$, scales with the corresponding eccentricity, $\epsilon_2$. Flow can as such be understood of a response to an initial geometric configuration -- the same is true for the string shoving model. In a heavy ion collision such a geometry is usually defined through the Glauber model, possibly plus fluctuations. In small systems, \pA and \pp collisions, this is more challenging, as sub-nucleonic degrees of freedom dominate the spatial distribution. In order for the shoving model to produce reliable results, a reasonable model for the spatial distribution of strings must be introduced. In order to calculate the transverse distributions of gluons in the colliding nucleons, the Mueller dipole model \cite{Mueller:1993rr} in a new Monte Carlo implementation \cite{Bierlich:2019wld} is used. The basis of the calculation, is the (leading order) branching probability for a dipole $r_{12}$, spanned between $\vec{r}_2$ and $\vec{r}_2$ in transverse space, to branch to two dipoles $r_{13}$ and $r_{23}$, when evolved in rapidity ($y$):
\begin{equation}
	\label{eq:mueller}
	\frac{\mathrm{d}\mathcal{P}}{\mathrm{d}y} = \mathrm{d}^2 \vec{r}_3 \frac{N_c \alpha_s}{2\pi^2} \frac{r^2_{12}}{r^2_{13}r^2_{23}}.
\end{equation}
Given an initial state to evolve, this allows for the creation of projectile and target Fock states for individual collisions, by the means of a parton shower-like evolution, where equation (\ref{eq:mueller}) is modified by a Sudakov form factor:
\begin{equation*}
	\exp\left(-\frac{N_c \alpha_s}{2\pi^2}\int_{y_{min}}^y \mathrm{d}y \int\mathrm{d}^2\vec{r}_3 \frac{r_{12}^2}{r^2_{13}r^2_{23}}\right).
\end{equation*}
This allows for an evolution strategy (``winner takes it all'') where a cascade is ordered in rapidity, with the possible emission in which the lowest rapidity is chosen in each step of the evolution. This process can be iterated until no emissions are allowed without breaking a maximal
rapidity, governed by the collision energy. The interaction between projectile and target can then be factorized into dipole--dipole interactions (between dipole $r_{12}$ and $r_{34}$ from projectile and target respectively), given by:
\begin{equation}
	\frac{\mathrm{d}\sigma_{dip}}{\mathrm{d}^2\vec{b}} = \frac{\alpha^2_s}{2} \log^2\left(\frac{r_{13}r_{24}}{r_{14}r_{23}}\right) \equiv f_{ij}.
\end{equation}
Assuming that all dipole--dipole interactions are independent of each other, multiple scatterings exponentiate, resulting in a unitarized scattering amplitude for a single collision:
\begin{equation}
	T(\vec{b}) = 1 - \exp\left(-\sum_{ij} f_{ij}\right).
\end{equation}
The scattering amplitude can be related to cross sections through the optical theorem, fixing all parameters\footnote{A proton initial state before evolution needs to be modelled, and a model for confinement is also introduced, giving a total of four parameters of the model (including $\Lambda_{QCD}$).}, leaving calculations of the proton substructure as real predictions. Interestingly, studies of eccentricities of \pp collisions reveals that a full dipole treatment cannot be distinguished from a simple ansatz, where MPIs are distributed according to a 2D Gaussian -- but studies of \pA collisions can. In figure \ref{fig:dipole-struc} (left), normalized symmetric cumulants as a function of final state charged multiplicity in \pp, \pA and \AA are shown. Normalized symmetric cumulants signifies the normalized correlation between flow coefficients, and therefore provides a picture of the initial state, given the aforementioned proportionality between eccentricities and flow coefficients, as:
\begin{align}
	\mathrm{NSC}(n,m)=&\frac{\langle v_n^2v_m^2\rangle-\langle v_n^2\rangle\langle
v_m^2\rangle}{\langle v_n^2\rangle\langle v_m^2\rangle}
\approx\frac{\langle \epsilon_n^2\epsilon_m^2\rangle-\langle \epsilon_n^2\rangle\langle
\epsilon_m^2\rangle}{\langle
\epsilon_n^2\rangle\langle\epsilon_m^2\rangle}.
\end{align}
Only for \pA collisions can the models be qualitatively distinguished, and interestingly, the dipole treatment is the only one predicting a negative NSC$(2,3)$ in line with ALICE data \cite{Acharya:2019vdf}, as opposed to the Gaussian or a normal Glauber treatment. The particular influence on \pA collisions can also be seen from the slightly improved description of flow fluctuation measured by CMS \cite{Sirunyan:2019pbr}, shown in figure \ref{fig:dipole-struc} (right), where the ratio is similarly expected to make the response coefficients cancel each other.

\section*{Acknowledgements}
I thank the organisers for a pleasant conference. Support from Swedish Research Council, contract number 2017-003, is gratefully acknowledged.
\bibstyle{woc}
\bibliography{ismd}

\begin{thebibliography}{29}

\bibitem{Sjostrand:2014zea}
T.~Sjöstrand, S.~Ask, J.R. Christiansen, R.~Corke, N.~Desai, P.~Ilten,
  S.~Mrenna, S.~Prestel, C.O. Rasmussen, P.Z. Skands, Comput. Phys. Commun.
  \textbf{191}, 159 (2015), \texttt{1410.3012}

\bibitem{Sjostrand:1987su}
T.~Sjostrand, M.~van Zijl, Phys. Rev. \textbf{D36}, 2019 (1987)

\bibitem{Ortiz:2013yxa}
A.~Ortiz~Velasquez, P.~Christiansen, E.~Cuautle~Flores, I.~Maldonado~Cervantes,
  G.~Paić, Phys. Rev. Lett. \textbf{111}, 042001 (2013), \texttt{1303.6326}

\bibitem{Bierlich:2015rha}
C.~Bierlich, J.R. Christiansen, Phys. Rev. \textbf{D92}, 094010 (2015),
  \texttt{1507.02091}

\bibitem{Bierlich:2018lbp}
C.~Bierlich, Nucl. Phys. \textbf{A982}, 499 (2019), \texttt{1807.05271}

\bibitem{Bierlich:2016smv}
C.~Bierlich, G.~Gustafson, L.~Lönnblad, JHEP \textbf{10}, 139 (2016),
  \texttt{1607.04434}

\bibitem{Bierlich:2018xfw}
C.~Bierlich, G.~Gustafson, L.~Lönnblad, H.~Shah, JHEP \textbf{10}, 134 (2018),
  \texttt{1806.10820}

\bibitem{Bialas:1976ed}
A.~Bialas, M.~Bleszynski, W.~Czyz, Nucl. Phys. \textbf{B111}, 461 (1976)

\bibitem{Andersson:1986gw}
B.~Andersson, G.~Gustafson, B.~Nilsson-Almqvist, Nucl. Phys. \textbf{B281}, 289
  (1987)

\bibitem{Bierlich:2019rhm}
C.~Bierlich et~al., SciPost Phys. \textbf{8}, 026 (2020), \texttt{1912.05451}

\bibitem{Bierlich:2020wms}
C.~Bierlich et~al. (2020), \texttt{2001.10737}

\bibitem{Andersson:1983ia}
B.~Andersson, G.~Gustafson, G.~Ingelman, T.~Sjostrand, Phys. Rept. \textbf{97},
  31 (1983)

\bibitem{Andersson:1983jt}
B.~Andersson, G.~Gustafson, B.~Soderberg, Z. Phys. \textbf{C20}, 317 (1983)

\bibitem{Sjostrand:1984ic}
T.~Sjostrand, Nucl. Phys. \textbf{B248}, 469 (1984)

\bibitem{Abramovsky:1988zh}
V.A. Abramovsky, E.V. Gedalin, E.G. Gurvich, O.V. Kancheli, JETP Lett.
  \textbf{47}, 337 (1988), [Pisma Zh. Eksp. Teor. Fiz.47,281(1988)]

\bibitem{Bierlich:2017vhg}
C.~Bierlich, G.~Gustafson, L.~Lönnblad, Phys. Lett. \textbf{B779}, 58 (2018),
  \texttt{1710.09725}

\bibitem{Biro:1984cf}
T.S. Biro, H.B. Nielsen, J.~Knoll, Nucl. Phys. \textbf{B245}, 449 (1984)

\bibitem{Bierlich:2014xba}
C.~Bierlich, G.~Gustafson, L.~Lönnblad, A.~Tarasov, JHEP \textbf{03}, 148
  (2015), \texttt{1412.6259}

\bibitem{Baker:1991bc}
M.~Baker, J.S. Ball, F.~Zachariasen, Phys. Rept. \textbf{209}, 73 (1991)

\bibitem{Cea:2014uja}
P.~Cea, L.~Cosmai, F.~Cuteri, A.~Papa, Phys. Rev. \textbf{D89}, 094505 (2014),
  \texttt{1404.1172}

\bibitem{Khachatryan:2010gv}
V.~Khachatryan et~al. (CMS), JHEP \textbf{09}, 091 (2010), \texttt{1009.4122}

\bibitem{Bierlich:2019ixq}
C.~Bierlich, Phys. Lett. \textbf{B795}, 194 (2019), \texttt{1901.07447}

\bibitem{ATLAS:2017nkt}
T.A. collaboration (ATLAS) (2017)

\bibitem{Cacciari:2008gp}
M.~Cacciari, G.P. Salam, G.~Soyez, JHEP \textbf{04}, 063 (2008),
  \texttt{0802.1189}

\bibitem{Chatrchyan:2013vbb}
S.~Chatrchyan et~al. (CMS), JHEP \textbf{05}, 090 (2013), \texttt{1303.4811}

\bibitem{Mueller:1993rr}
A.H. Mueller, Nucl. Phys. \textbf{B415}, 373 (1994)

\bibitem{Bierlich:2019wld}
C.~Bierlich, C.O. Rasmussen, JHEP \textbf{10}, 026 (2019), \texttt{1907.12871}

\bibitem{Acharya:2019vdf}
S.~Acharya et~al. (ALICE), Phys. Rev. Lett. \textbf{123}, 142301 (2019),
  \texttt{1903.01790}

\bibitem{Sirunyan:2019pbr}
A.M. Sirunyan et~al. (CMS), Phys. Rev. \textbf{C101}, 014912 (2020),
  \texttt{1904.11519}

\end{thebibliography}

\end{document}